\documentclass[final]{elsarticle}

\usepackage{lineno,hyperref}
\usepackage{multirow}
\usepackage{graphicx}

\usepackage{tabularx}
\makeatletter
\def\hlinewd#1{%
\noalign{\ifnum0=`}\fi\hrule \@height #1 %
\futurelet\reserved@a\@xhline}
\makeatother

\journal{Journal of Alloys and Compounds}

\begin{document}

\begin{frontmatter}

\title{Thermal conductivity of rare-earth scandates in comparison to other oxidic substrate crystals}

\author[mainaddress,secondaryaddress]{Julia Hidde}
\author[mainaddress]{Christo Guguschev}
\author[mainaddress]{Steffen Ganschow}

\author[mainaddress,secondaryaddress]{Detlef Klimm\corref{correspondingauthor}}
\cortext[correspondingauthor]{Corresponding author}
\ead{detlef.klimm@ikz-berlin.de}

\address[mainaddress]{Leibniz-Institut f\"ur Kristallz\"uchtung, Max-Born-Str. 2, 12489 Berlin, Germany}
\address[secondaryaddress]{Humboldt-Universit\"at zu Berlin, Institut f\"ur Chemie, Brook-Taylor-Str. 2, 12489 Berlin}

\begin{abstract} 
High-temperature thermal properties of three neighboring rare-earth scandates DyScO$_3$, TbScO$_3$ and GdScO$_3$ were compared to La$_{0.29}$Sr$_{0.71}$Al$_{0.65}$Ta$_{0.35}$O$_3$ (LSAT) and sapphire. To calculate thermal conductivity, heat capacity and thermal diffusivity were measured by differential scanning calorimetry and laser flash technique, respectively. DyScO$_3$ and TbScO$_3$ showed an untypical rise in the thermal conductivity above 900\,K, while for GdScO$_3$, LSAT and sapphire the expected decrease at elevated temperatures could be observed. These results lead to the proposal of a new type of heat transport by migrating ions.
\end{abstract}


\end{frontmatter}


\section{Introduction}

Over the last few decades tremendous advances have been made in the development of digital electronic devices, which are widely used in human life. The central element of these devices is the processing unit, which in turn consists of millions of transistors. For these devices the trend towards miniaturization remains unbroken. However, the technological evolution has reached a limit regarding the scaling of SiO$_2$ in metal-oxide semiconductor field-effect transistors (MOSFETs). Below a certain thickness, SiO$_2$ is no longer an effective gate insulator \cite{Schlom04}. This problem can be overcome by replacing SiO$_2$ with a ``high-$\kappa$'' gate dielectric, which has a higher dielectric constant. Besides HfO$_2$, rare-earth scandates have been shown to be among the promising candidates for this application \cite{Schlom02,Zhao05, Wagner06, Christen06, Ozben11, Pampillon17}.

These compounds with the general formula REScO$_3$ where reported as congruently melting for the larger RE$^{3+}$ from La$^{3+}$ (octahedral radius 117.2\,pm) down to Dy$^{3+}$ (105.2\,pm). All REScO$_3$ within this series crystallize as distorted perovskites in an orthorhombic structure (space group $Pnma$) with pseudocubic lattice constants around 4\,\AA\ \cite{Clark78,Velickov07,Velickov08}. This makes them also interesting as substrate materials for the epitaxy of other perovskites that are e.g. superconducting, ferroelectric, or multiferroic \cite{Haeni04}.

Mateika et al. \cite{Mateika91} proposed several other substrate materials which are mixed crystals; among them the cubic perovskite La$_{0.29}$Sr$_{0.71}$Al$_{0.65}$Ta$_{0.35}$O$_3$ (LSAT) became most relevant and is offered commercially meanwhile. Different groups are growing this material with slightly different chemical composition, and with consequently somewhat different lattice constant. Nowadays, LSAT is also considered as prospective substrate material for the epitaxial deposition of GaN layers \cite{Sakowska01,Shimamura98}.

Besides the somewhat ``exotic'' REScO$_3$ and LSAT crystals, sapphire wafers (better to say corundum, $\alpha$-Al$_2$O$_3$) are now the technological basis for the production of nitride light emitting diodes that are used for solid state lighting \cite{Nakamura94}. Moreover, sapphire crystals are used in scales of tons for windows and watch displays.

During the corresponding crystal growth processes for the production of these materials a sufficient amount of heat must be flowing from the hot melt through the growth interface to the colder seed. The adjustment of these heat flows is not always straightforward, and especially for materials with low optical transmissivity in the near infrared, in combination with low thermal conductivity $\lambda$, growth instabilities or even growth disruption can occur. Especially for some REScO$_3$ the formation of ``growth spirals'' is a typical issue \cite{Uecker06b}.

Another issue is connected with the operation of nanoscale semiconductor devices. There an increased heat generation occurs due to a significant reduction of the dynamic power, whereas the power dissipation per unit area remains constant \cite{Figena08}. Therefore, the heat dissipation characteristics of these devices are of great importance and need to be studied in detail. In this context thermal conductivity is a relevant variable, in order to analyze and model the self-heating behavior of MOSFETs and adjust the device architecture \cite{Figena08, Berger91, Ghazanfarian12}.

The current study investigates the thermal conductivity of three different rare-earth scandates DyScO$_3$, TbScO$_3$ and GdScO$_3$ in comparison with two widely used compounds La$_{0.29}$Sr$_{0.71}$Al$_{0.65}$Ta$_{0.35}$O$_3$ (LSAT) and Al$_2$O$_3$ (corundum, sapphire). The consideration of sapphire seemed useful because $\lambda$ data given by several crystal producers are significantly different: For the $[100]$ and $[001]$ directions one finds e.g. 23.0 and 25.8\,W/(m$\cdot$K) at room temperature \cite{sapp-valley}, whereas another reference reports 42\,W/(m$\cdot$K) without specification of the crystallographic direction \cite{sapp-kyocera}. Other references report a larger conductivity in $[100]$ than in $[001]$ \cite{sapp-isi}. Data for elevated temperatures are extremely scarce (but see \cite{sapp-kyocera}). This paper will report and compare experimental data for the crystals mentioned above. All single crystals were grown in the author's laboratory under similar experimental conditions, and also the $\lambda(T)$ measurements were obtained under identical conditions for all samples.

\section{Theoretical background}
\label{sec:background}

When a temperature gradient is applied to a solid object, thermal energy will flow from the warmer to the cooler zone. The thermal conductivity $\stackrel{2\rightarrow}{\lambda}$ is a material property, which describes the rate of heat transport in the object relating the local heat flux $\vec{J}$ to the temperature gradient $\partial T/ \partial x\equiv\vec{\nabla} T$ \cite{Encyclopedia, AnisotropieBuch}. 

\begin{equation}
\vec{J}=-\stackrel{2\rightarrow}{\lambda} \cdot \vec{\nabla} T      \label{Eq_allg}
\end{equation}

Like for other transport properties, $\stackrel{2\rightarrow}{\lambda}$ is a symmetric second rank tensor, which defines the relation between an applied and resultant vector. Thus, for an anisotropic material it consists of a maximum of six independent components. Though this number can be reduced by consideration the crystal symmetry \cite{AnisotropieBuch}. For the different crystal systems one obtains the following numbers of independent components that are given in brackets: triclinic (6); monoclinic (4); orthorhombic (3); trigonal, tetragonal, hexagonal (2); cubic (1) \cite{MineralsBuch}. This means that only in cubic, polycrystalline or other isotrope materials the thermal conductivity is described by one number $\lambda$.

All rare-earth scandates are orthorhombic perovskites \cite{Liferovich04}. Sapphire ($\alpha$-Al$_2$O$_3$) has a trigonal structure \cite{Dashevsky04} and LSAT is cubic \cite{Chakoumakos98}. This means that only three or less components of thermal conductivity need to be measured in order to describe the corresponding $\stackrel{2\rightarrow}{\lambda}$ tensors.

The thermal conductivity of a given material can be derived from the thermal diffusivity $a$, because these two properties are linked by 
\begin{equation}
\lambda=\varrho\cdot c_p \cdot a              \label{Eq_exp}
\end{equation}
where $c_p$ is the specific heat capacity and $\varrho$ the mass density \cite{Parker61}.

Equation (\ref{Eq_allg}) describes only the heat that is transported through the solid via phonons, electrons, or any other carriers. In transparent media, an additional share of heat energy is transported by radiation. Fortunately, both mechanisms work on very different time scales, because heat radiation moves with the speed of light; phonons in contrast move with the speed of sound. In the laboratory scale for distances around 1\,mm this means that heat radiation arrives almost immediately, whereas conducted heat (\ref{Eq_allg}) arrives after ca. 1\,ms, which can be separated well \cite{Mehling98}.

\section{Experimental} 

The crystals investigated in this study were grown by the conventional Czochralski technique from a cylindrical iridium crucible and in a growth atmosphere of flowing argon. For the growth of LSAT, typically 10\% CO$_2$ were added to Ar. The pulling rates were typically 1.0--1.5\,mm/h and the rotation rates between 8 and 15\,rpm \cite{Shimamura98,Uecker05,Uecker08}. The thermal conductivity as a function of temperature was determined according to equation (\ref{Eq_exp}). For this purpose, temperature dependent measurements of thermal transport and heat capacity were performed on the samples, while the mass density $\varrho$ was taken from the literature \cite{Velickov07, Velickov08, Uecker08, Ito02, Lucasiewicz02, Yim73, Slack62}. Table~\ref{Tab_Dichten} lists the data used for $\varrho$ near room temperature. The average linear expansion coefficient $\overline{\alpha}$ was used to calculate the temperature dependence of the density and the thickness by the change of the sample dimensions. It should be mentioned, however, that the relative change of thickness upon heating by $\Delta T\approx1000$\,K is only $\overline{\alpha}\cdot\Delta T$, which is in the order of 1\%. Correspondingly the influence on $\varrho$ is in the order of $3\cdot\overline{\alpha}\cdot\Delta T$, which is 3\%.

\begin{table}[ht]
\centering
\caption{Mass density at room temperature $\varrho$ and average linear expansion coefficient $\overline{\alpha}$ of the investigated compounds.}
\begin{tabular}{lll}
\hline
Crystal     & $\varrho$ (g/cm$^3$)        & $\overline{\alpha}$ ($10^{-6}$\,K$^{-1}$)  \rule{0pt}{10pt}\\ \hline
DyScO$_3$   & 6.90 \cite{Velickov07}      & 8.4 \cite{Uecker08}  \\ 
TbScO$_3$   & 6.55 \cite{Velickov08}      & 8.2 \cite{Uecker08}  \\ 
GdScO$_3$   & 6.66 \cite{Velickov07}      & 10.9 \cite{Uecker08} \\ 
LSAT        & 6.65 \cite{Lucasiewicz02}   & 9.9 \cite{Ito02}     \\ 
sapphire    & 3.99 \cite{Slack62}         & 7.7 \cite{Yim73}     \\ \hline
\end{tabular}
\label{Tab_Dichten}
\end{table}

The specific heat capacity $c_p$ was measured by heat flux differential scanning calorimetry (DSC) with a NETZSCH STA449C. For this purpose, three DSC measurements with linear heating ramps were performed: first with an empty crucible as baseline correction, second with a reference substance where $c_p(T)$ is known and third with the sample to analyze \cite{Neill66}. The heat capacity was then calculated by comparison of the obtained curves using the DIN 51007 method.

All DSC measurements were carried out in Pt crucibles with lids in flowing argon. The crystal samples were pestled to crush and Al$_2$O$_3$ powder was used as reference. To ensure reproducibility, four subsequent heating runs with 20\,K/min from 313\,K to 1473\,K were performed during every measurement. Because the first run showed higher experimental scatter and deviated by up to 15\% from the following runs, solely the average values of the subsequent runs were used to determine $c_p(T)$.

The thermal diffusivity $a$ was measured by laser flash technique with a NETZSCH LFA427. To determine $a(T)$ with this method the bottom side of the sample was irradiated by a Nd:YAG laser, which generated a heat pulse. Transport of that heat thought the material resulted in a rise of the temperature on the top of the sample, which was measured and evaluated \cite{Parker61}. For this purpose, the temperature time response was fitted by Mehling’s model \cite{Mehling98} for semi-transparent media to obtain $a(T)$. 

LFA measurements were performed on sample slices with thicknesses of approximately 1.5\,mm and diameters between 6 and 12\,mm, which were cut to give slice orientations (100), (010) or (001), respectively. To ensure uniform heat absorption, the samples where covered with a thin layer of graphite prior to every measurement. During the measurements a protective atmosphere of flowing nitrogen prevented oxidation of that graphite layer. The temperature was risen stepwise from 298\,K to typically 1473\,K, with a typical difference of 50\,K between the steps. Degradation of the graphite layer resulted in a lower temperature limit for some samples only. At every temperature a series of three laser shots was performed using a laser voltage of 460\,V and a pulse width of 0.8\,ms, which results in a laser shot energy of approximately 4\,J. The thermal diffusivity was then calculated as the average of two to five LFA measurements.

\section{Results and Discussion}

DSC measurements of DyScO$_3$, TbScO$_3$, GdScO$_3$ and LSAT powder samples revealed a smooth and monotonously rising behavior for the specific heat capacity $c_p(T)$ from room temperature to 1473\,K. As expected, the heat capacity of the investigated compounds showed no spikes, indicating that no phase transition occurred over the mentioned temperature range. Reference data for $c_p(T)$ of Al$_2$O$_3$ were taken from the literature \cite{FactSage}.

It should be noted, that DSC measurements with single crystalline samples resulted in less reliable DSC curves than with powders, probably resulting from low thermal conductivity of most samples (see below), which can lead to an inhomogeneous temperature distribution in the crystalline sample during DSC measurements \cite{Schenker97}. In that case, the large number of grains in a powder can solve the problem by providing an isotropic thermal conductivity of the measured samples. Recently, analogous observations were reported for CaSc$_2$O$_4$ \cite{Guguschev17}.

The experimental data for $c_p(T)$ of the powder samples could be fitted reasonably by a polynom of type (\ref{Eq_cp}). Fit parameter $a$, $b$ and $c$ for the investigated compounds are given in Table~\ref{Tab_cp}.

\begin{equation}
c_p = a + b\cdot T + \frac{c}{T^2}  \label{Eq_cp}
\end{equation}

\begin{table}[ht]
\centering
\caption{Fit parameters and coefficient of determination $R^2$ for the $c_p(T)$ functions (\ref{Eq_cp}) of the investigated compounds.}
\begin{tabular}{l|ccccc}
\hline
\multicolumn{1}{c|}{\multirow{2}{*}{Compound}} & \multicolumn{4}{c}{Parameter }               \\ \cline{2-5} 
\multicolumn{1}{c|}{}  & $a$      & $b$          & $c$            & $R^2$\rule{0pt}{10pt}      \\ \hline
DyScO$_3$\rule{0pt}{10pt}  & $116.103$ & $6.3179 \cdot 10^{-3}$ & $-2.39870 \cdot 10^{6}$ & $0.9422$ \\ 
TbScO$_3$  & $97.6047$ & $1.9098 \cdot 10^{-2}$ & $-1.45446 \cdot 10^{6}$ & $0.9521$ \\ 
GdScO$_3$  & $98.3548$ & $1.0983 \cdot 10^{-2}$ & $-1.56790 \cdot 10^{6}$ & $0.7800$ \\ 
LSAT       & $101.568$ & $1.1860 \cdot 10^{-2}$ & $-2.26236 \cdot 10^{6}$ & $0.9372$ \\ \hline
\end{tabular}
\label{Tab_cp}
\end{table}

\begin{figure}[htb]
\centering
\includegraphics[width=0.5\textwidth]{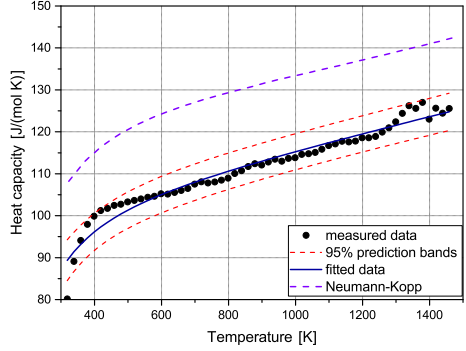}
\caption{Measured $c_p(T)$ data for TbScO$_3$, together with a fit to the function in equation (\ref{Eq_cp}) and 95\% prediction bands. For comparison $c_p(T)$ data calculated according to the Neumann-Kopp rule are also shown.}
\label{Abb_Tb_cp}
\end{figure}

The Neumann-Kopp rule allows a theoretical approximation of $c_p(T)$ from the weighted sum of heat capacities of their constituent binary oxides \cite{Aubel1901}. Experimental data of all investigated materials correlate satisfactorily with the theoretical estimation, showing only a slight deviation of about 10--15\%. The lower experimental values are probably attributed to the reaction energy for the formation of each respective oxide, which is not taken into account by the Neumann-Kopp rule. As an example, Figure~\ref{Abb_Tb_cp} shows experimental data for TbScO$_3$ that are fitted to equation (\ref{Eq_cp}). The theoretical approximation by application of the Neumann-Kopp rule is added for comparison. For a representation of $c_p(T)$ for DyScO$_3$, GdScO$_3$ and LSAT see supplementary Figures\,S1--S3. 

Thermal conductivity $\lambda(T)$ of the samples was calculated from the measured thermal diffusivity $a(T)$ with equation (\ref{Eq_exp}) using the fitted experimental powder data for $c_p(T)$. As the low variation of the heat capacity at elevated temperatures allows a similar description of the temperature dependence of $\lambda(T)$ and $a(T)$, only $\lambda(T)$ will be discussed in this study. Values for $a(T)$ are presented in Table~S1--S2 in the supplementary Material.

\begin{figure}[htb]
\centering
\includegraphics[width=0.9\textwidth]{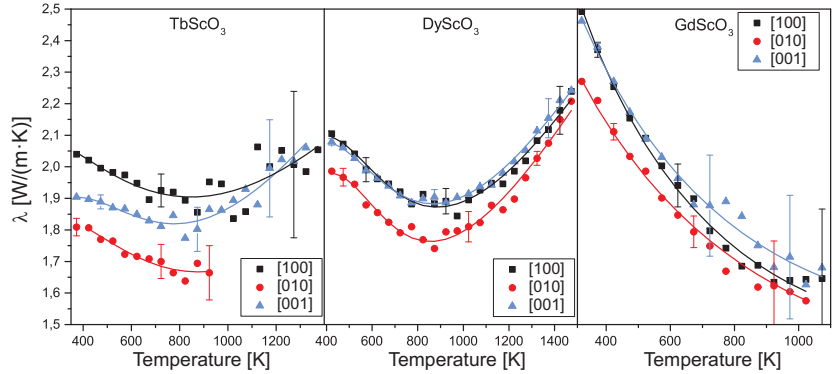}
\caption{Thermal conductivity data of the three investigated rare-earth-scandates fitted with the functions given in table~\ref{Tab_Warmeleit}. The error bars indicate the doubled standard deviation.}
\label{Abb_Warmel_Scandate}
\end{figure}

Figure~\ref{Abb_Warmel_Scandate} presents the thermal conductivity of the investigated rare-earth scandates. All samples show $\lambda$ values in the range 1.5--2.5\,W/(m$\cdot$K), which is significantly lower than for most other functional oxides. Besides, the temperature dependence $\lambda(T)$ is not very strong. In particular DyScO$_3$ and TbScO$_3$ revealed a marginal variation of the thermal conductivity of only about 20\% over the entire investigated temperature range. But the most remarkable fact is the increase of $\lambda(T)$ for these two rare-earth scandates at temperatures above 900\,K, while $\lambda(T)$ of GdScO$_3$ drops monotonously. Recently, cross-plane $\lambda(T)$ data below room temperature ($20\,\mathrm{K}\leq T\leq350\,\mathrm{K}$) of several commercial REScO$_3$ substrates with (110) orientation, including DyScO$_3$, TbScO$_3$, and GdScO$_3$, were published by Langenberg et al. \cite{Langenberg16}. At 350\,K values of 2.8, 2.9, and 3.5\,W/(m$\cdot$K) were found in this order. One can expect that the $3\omega$ method that was used by Langenberg et al. results in values that are more influenced by the maximum $\stackrel{2\rightarrow}{\lambda}$ tensor component, which corresponds to the [100] direction. At 350\,K one reads from Figure~\ref{Abb_Warmel_Scandate} for [100] $\lambda=2.1$, $2.05$, and 2.5\,W/(m$\cdot$K), respectively, for these compounds. This is a similar trend, but the current values are ca. 25--30\% smaller, compared to reference \cite{Langenberg16}.

For many isolators thermal conductivity is dominated by phonons. At higher temperatures this typically results in a decrease of $\lambda$, because scattering reduces the phonon mean free path \cite{Slack62}. Euken's empirical law where thermal conductivity behaves inversely proportional to the temperature has been discovered experimentally in 1911 and holds for many materials \cite{Eucken1911}. This $\lambda\propto1/T$ law could be reproduced by theoretical consideration of three-phonon scattering events by Peierls 1929 \cite{Peierls1929}. However, if four-phonon processes are taken into account, higher exponents are needed to describe the temperature dependence. Hofmeister et al. proposed formulas in the form of equation\,(\ref{Eq_Warmel1}) and\,(\ref{Eq_Warmel2}) where $a,b$ and $c$ are constants \cite{Hofmeister06}.

\begin{equation}
\lambda = a + \frac{b}{T} + \frac{c}{T^2}  \label{Eq_Warmel1}
\end{equation}

\begin{equation}
\lambda = \frac{1}{a + b \cdot T + c \cdot T^2}  \label{Eq_Warmel2}
\end{equation}

\begin{figure}[htb]
\centering
\includegraphics[width=0.95\textwidth]{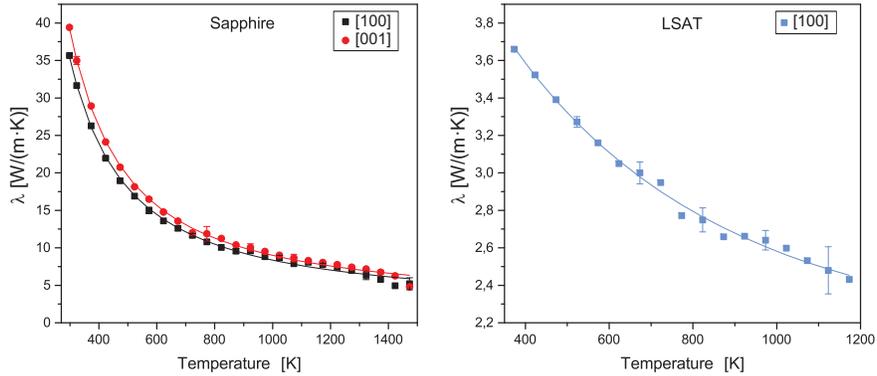}
\caption{Thermal conductivity data of sapphire and LSAT fitted with the functions given in Table~\ref{Tab_Warmeleit}. The error bars indicate the doubled standard deviation.}
\label{Abb_Warmel_LSAT_Sapp}
\end{figure}

Thermal conductivity of the investigated samples could be fitted to $\lambda=a+b/T$ only over a limited temperature range between 400 and 800\,K. Hence, the $1/T$ law is insufficient to describe thermal conductivity of the studied compounds. For GdScO$_3$ and LSAT (see Figure~\ref{Abb_Warmel_LSAT_Sapp}), the experimental data could be represented satisfactorily by equation~(\ref{Eq_Warmel2}), while for sapphire a better fit was obtained with equation\,(\ref{Eq_Warmel1}). However, both models fail to explain the growing $\lambda(T)$ of DyScO$_3$ and TbScO$_3$ for $T>900$\,K.

Recently, an analogous increase of the thermal conductivity could be observed for the inverse spinel MgGa$_2$O$_4$ \cite{Schwarz15} as well as the pyrochlore Tb$_2$Ti$_2$O$_7$ \cite{Klimm17}. For the spinel it was suggested, that cations exchanging their site at elevated temperatures contribute to thermal transport. Thus, it seems plausible to consider transport of thermal energy in isolators through moving ions. An empirical formula with a linear term added to equation\,(\ref{Eq_Warmel2}) can describe $\lambda(T)$ well (see equation\,(\ref{Eq_Warmel3})). Here the first inverse term describes the heat transport by phonons, while the second linear term might be related to transportation of thermal energy by any type of charge carriers \cite{Klimm17}. For electrically conducting materials, free electrons could contribute to heat transport \cite{Glasbrenner63}. However, this mechanism seems not realistic in the present case, because the electron density of the investigated rare-earth scandates is low \cite{Zhao05, Ozben11}. At least for the REScO$_3$ a significant portion (ca. 4\% for TbScO$_3$ \cite{Velickov08}) of the RE$^{3+}$ sites is empty, which could enable diffusive movements of ions.

\begin{equation}
\lambda = \frac{1}{a + b \cdot T + c \cdot T\textsuperscript{2}} + d \cdot T \label{Eq_Warmel3}
\end{equation}

\begin{table}[ht]
\centering
\caption{Formula and fit parameters for the $\lambda(T)$ functions of the investigated crystals.}
\begin{tabular}{  l | ccccc}
\hline
\multirow{2}{*}{Crystal} & \multicolumn{5}{c}{Parameter}                           \\ \cline{2-6} 
                         & $a$  & $b$  & $c$    & $d$        & $R^2$\rule{0pt}{10pt}     \\ \hline
\multicolumn{6}{c}{$\lambda = 1/(a+bT+cT^2) + dT$} \rule{0pt}{10pt}                                    \\ \hline
DyScO$_3$ [100]   & $0.7598$ & $-1.1801 \cdot 10^{-3}$ & $2.1802 \cdot 10^{-6}$  & $1.3213 \cdot 10^{-3}$ & $0.9730$  \\ 
DyScO$_3$ [010]   & $1.0328$ & $-2.2087 \cdot 10^{-3}$ & $3.4104 \cdot 10^{-6}$  & $1.3477 \cdot 10^{-3}$ & $0.9757$ \\ 
DyScO$_3$ [001]   & $0.7481$ & $-1.0967 \cdot 10^{-3}$ & $2.1126 \cdot 10^{-6}$  & $1.3388 \cdot 10^{-3}$ & $0.9911$ \\ 
TbScO$_3$ [100]   & $0.4734$ &  $1.4537 \cdot 10^{-4}$ & $6.0391 \cdot 10^{-7}$  & $1.1017 \cdot 10^{-3}$ & $0.5583$ \\ 
TbScO$_3$ [010]   & $0.5779$ & $-6.8396 \cdot 10^{-5}$ & $1.0022 \cdot 10^{-6}$  & $1.0157 \cdot 10^{-3}$ & $0.9037$ \\ 
TbScO$_3$ [001]   & $0.6941$ & $-5.9094 \cdot 10^{-4}$ & $1.6570 \cdot 10^{-6}$  & $1.2996 \cdot 10^{-3}$ & $0.9075$ \\ 
\hline
\multicolumn{6}{c}{$\lambda = 1/(a+bT+cT^2)$}  \rule{0pt}{10pt}                                        \\ \hline
GdScO$_3$ [100]   & $0.2308$ & $5.8613 \cdot 10^{-4}$  & $-1.9817 \cdot 10^{-7}$ & -          & $0.9939$ \\ 
GdScO$_3$ [010]   & $0.3072$ & $4.5118 \cdot 10^{-4}$  & $-1.2866 \cdot 10^{-7}$ & -          & $0.9755$ \\
GdScO$_3$ [001]   & $0.2639$ & $4.8709 \cdot 10^{-4}$  & $-1.5736 \cdot 10^{-7}$ & -          & $0.9857$ \\ 
LSAT                           & $0.1733$ & $2.9727 \cdot 10^{-4}$  & $-8.3220 \cdot 10^{-8}$ & -          & $0.9947$ \\ 
\hline
\multicolumn{6}{c}{$\lambda = a + b/T + c/T^2$}  \rule{0pt}{10pt}                                      \\ \hline
Sapphire [100]   & $1.6402$ & $5284.65$  & $1.4327 \cdot 10^{6}$  & - & $0.9976$ \\ 
Sapphire [001]   & $1.9450$ & $5303.74$  & $1.7424 \cdot 10^{6}$  & - & $0.9981$ \\ \hline
\end{tabular}
\label{Tab_Warmeleit}
\end{table}

Equation~(\ref{Eq_Warmel3}) approximates the experimental data for DyScO$_3$ and TbScO$_3$ well. Fit Parameters for all investigated compounds are given in Table~\ref{Tab_Warmeleit} together with the formula, which was found to give the best description of $\lambda(T)$. In Figures~\ref{Abb_Warmel_Scandate} and \ref{Abb_Warmel_LSAT_Sapp} the experimental data for thermal conductivity are shown together with the corresponding fit functions.

Since the rare-earth scandates are orthorhombic, three independent components are needed to describe the anisotropic thermal conductivity tensor (cf. section~\ref{sec:background}). All samples revealed a relatively low anisotropy, exceeding not more than fifteen percent as the biggest difference between the orientations. For TbScO$_3$ the [100] direction was found to have the highest thermal conductivity followed by [001] as the medium, and [010] showing the lowest values for $\lambda$. At 400\,K the difference between each of these orientations is about 0.1\,W/(m$\cdot$K). At elevated temperatures [100] and [001] seem to become isotropic, but that observation might be inaccurate because of the high experimental scatter. 

DyScO$_3$ shows at lower temperatures a similar trend $\lambda^{[100]}>\lambda^{[001]}>\lambda^{[010]}$. However, the anisotropy of the different directions is less pronounced, showing a distance of 0.08\,W/(m$\cdot$K) at 400\,K between [001] and [010], and only 0.03\,W/(m$\cdot$K) between [100] and [001]. For $T>600$\,K $\lambda^{[001]}$ becomes larger than $\lambda^{[100]}$, while $\lambda^{[010]}$ remains significantly lower.

For $T>400$\,K GdScO$_3$ has its highest thermal conductivity along [001], which is closely followed by $\lambda^{[100]}$. However, the anisotropy is small even at room temperature. As for the other scandates, $\lambda^{[010]}$ is the smallest value, and at 400\,K $\lambda^{[100]}$ is by ca. 0.15\,W/(m$\cdot$K) larger than $\lambda^{[010]}$. 

\begin{figure}[htb]
\centering
\includegraphics[width=0.95\textwidth]{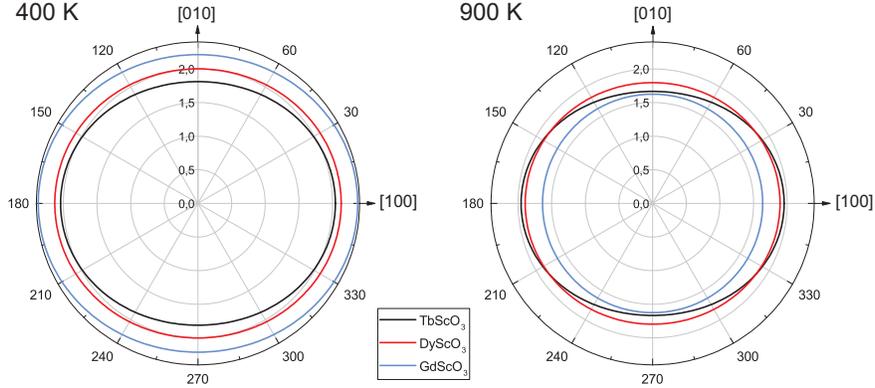}
\caption{Cross section through the representation surface of the thermal conductivity tensor of the investigated rare-earth scandates along the $[100] - [010]$ plane, the component of the [001] orientation is not shown. Radial units are W/(m$\cdot$K).}
\label{Abb_Polar}
\end{figure}

Figure~\ref{Abb_Polar} illustrates the anisotropy of the rare-earth scandates as a cross section through the representation surface of the thermal conductivity tensor perpendicular [001], and the anisotropies are shown as ellipses. At 400\,K (Figure~\ref{Abb_Polar}, left) $\lambda^{[100]}$ of all three scandates is noticeably larger than $\lambda^{[010]}$, which is represented by the elongated shape of the ellipse in that direction. At 900\,K (Figure~\ref{Abb_Polar}, right) the anisotropy of TbScO$_3$ remains unchanged and is slightly reduced for DyScO$_3$, while GdScO$_3$ becomes almost isotropic. Of the three investigated rare-earth scandates, GdScO$_3$ shows the highest thermal conductivity at room temperature and the lowest conductivity at elevated temperatures.

Thermal conductivity of LSAT and sapphire is presented in Figure~\ref{Abb_Warmel_LSAT_Sapp}. These two compounds generally show a significantly higher thermal conductivity than the investigated rare-earth scandates. As LSAT is cubic, only one single value is needed to represent the thermal conductivity tensor, while it has two independent components for the trigonal sapphire. The LSAT value $4.42\pm0.27$\,W/(m$\cdot$K) given by Sakowska et al. \cite{Sakowska01} is slightly higher than our value, but these authors investigated a different composition of this solid solution crystal. Both LSAT and sapphire show the typical decrease of $\lambda$ with $T$. Of all investigated substrates, sapphire has the highest thermal conductivity, reaching values of 39.4\,W/(m$\cdot$K) for [001] and 35.6\,W/(m$\cdot$K) for [100] at room temperature. These values are slightly higher and in good agreement with the thermal diffusivity recently presented by Hofmeister \cite{Hofmeister14}.

Roufosse \& Klemens \cite{Roufosse73} derived that $\lambda$ should vary as the inverse cube root of the number of atoms in the primitive unit cell. This implies that basically a simple structure should conduct heat better than a more complicated one. Thus it is surprising that the ``complicated'' La$_{0.29}$Sr$_{0.71}$Al$_{0.65}$Ta$_{0.35}$O$_3$ shows significantly larger $\lambda$ for all $T$ than the ``simple'' REScO$_3$. Besides the $\lambda(T)$ minima around 700--900\,K that were found for TbScO$_3$ and DyScO$_3$ (Figure~\ref{Abb_Warmel_Scandate}) cannot be explained on a theory that relies solely on phonon scattering. These results, in connection with previous work of Schwarz et al. \cite{Schwarz15} for the inverse spinel MgGa$_2$O$_4$, lead to the assumption that ionic charge carriers might be involved in the transportation of thermal energy for some substances. 

\section{Conclusions}

Most of the investigated crystals (three rare-earth scandates REScO$_3$, ``LSAT'' La$_{0.29}$Sr$_{0.71}$Al$_{0.65}$Ta$_{0.35}$O$_3$) show low thermal conductivies below 4\,W/(m$\cdot$K) and a weak anisotropy in thermal transport. Only sapphire has a high thermal conductivity up to ca. 40\,W/(m$\cdot$K) at room temperature, which is already in the order of $\lambda=67$\,W/(m$\cdot$K) for polycrystalline iron that were determined by the inventors of the flash method which was used also in this study \cite{Slack62}. 

From the thermal diffusivity data for MgGa$_2$O$_4$ in reference \cite{Schwarz15} one calculates at 373\,K $\lambda\approx10$\,W/(m$\cdot$K) which is a rather high value, in between that of sapphire and LSAT (Figure~\ref{Abb_Warmel_LSAT_Sapp}). It is remarkable that all these highly conductive oxides are based on dense packings of spherical O$^{2-}$ ions, with hexagonal dense packing for sapphire and cubic dense packing for the perovskite (LSAT) and the spinel (MgGa$_2$O$_4$). The REScO$_3$ crystals, in contrast, are orthorhombically distorted perovskites, and hence not dense packed. One can imagine that dense packing of atoms or ions promotes the energy transport by phonons as the main mechanism of thermal conductivity.

\section*{Acknowledgements}

The authors gratefully thank M. Br\"utzam, I. Schulze-Jonack, M. Imming-Friedland, V. Lange, U. Juda, T. Wurche, and E. Thiede for material preparation and A. Kwasniewski for performing X-ray measurements for the determination of the crystal orientation.

\section*{References}


\end{document}